\documentclass[aps,prb,twocolumn,showpacs,10pt,groupedaddress]{revtex4-1}
\usepackage{amssymb}
\usepackage{graphicx}
\usepackage{amsfonts}
\usepackage{amsmath}
\usepackage[usenames]{color}

\begin{document}

%=================================================================================================
\title{Quasiparticle velocities in 2D electron/hole liquids with spin-orbit coupling}
%=================================================================================================
\author{D.~Aasen}
%\affiliation{Department of Physics, McGill University, Montr{\'e}al, Qu{\'e}bec H3A 2T8, Canada}
\author{Stefano Chesi}
%\affiliation{Department of Physics, McGill University, Montr{\'e}al, Qu{\'e}bec H3A 2T8, Canada}
\author{W.~A.~Coish}
\affiliation{Department of Physics, McGill University, Montr{\'e}al, Qu{\'e}bec H3A 2T8, Canada}

\date{\today}

\begin{abstract}
We study the influence of spin-orbit interactions on quasiparticle dispersions in two-dimensional electron and heavy-hole liquids in III-V semiconductors.  To obtain closed-form analytical results, we restrict ourselves to  spin-orbit interactions with isotropic spectrum and work within the screened Hartree-Fock approximation, valid in the high-density limit.  For electrons having a linear-in-momentum Rashba (or, equivalently, Dresselhaus) spin-orbit interaction, we show that the screened Hartree-Fock approximation recovers known results based on the random-phase approximation and we extend those results to higher order in the spin-orbit coupling. While the well-studied case of electrons leads only to a weak modification of quasiparticle properties in the presence of the linear-in-momentum spin-orbit interaction, we find two important distinctions for hole systems (with a leading nonlinear-in-momentum spin-orbit interaction).  First, the group velocities associated with the two hole-spin branches acquire a significant difference in the presence of spin-orbit interactions, allowing for the creation of spin-polarized wavepackets in zero magnetic field.  Second, we find that the interplay of Coulomb and spin-orbit interactions is significantly more important for holes than for electrons and can be probed through the quasiparticle group velocities. These effects should be directly observable in magnetotransport, Raman scattering, and femtosecond-resolved Faraday rotation measurements. Our results are in agreement with a general argument on the velocities, which we formulate for an arbitrary choice of the spin-orbit coupling.

\end{abstract}


%71.10.-w 	Theories and models of many-electron systems
%75.70.Tj 	Spin-orbit effects
%71.45.Gm 	Exchange, correlation, dielectric and magnetic response functions, plasmons 
%73.61.Ey 	III-V semiconductors 
%71.70.Ej 	Spin-orbit coupling, Zeeman and Stark splitting, Jahn-Teller effect 
%71.10.Ca 	Electron gas, Fermi gas 

\maketitle

%=======================================================================================
\section{Introduction}
%=======================================================================================

Semiconductor heterostructures offer the possibility of forming two-dimensional liquids with tunable sheet density $n_s$. In an idealized model where the carriers have parabolic dispersion with band mass $m$ and interact with Coulomb forces,\cite{TheBook} the only relevant quantity is the dimensionless Wigner-Seitz radius $r_s =1/\sqrt{\pi a_B^2 n_s}$, where $a_{B} = \hbar^2 \epsilon_{r}/me^2$ is the effective Bohr radius ($\epsilon_r$ is the dielectric constant). Since $r_s$ serves as the interaction parameter, changing $n_s$ allows for a systematic study of the effects of the Coulomb interaction. In particular, properties of quasiparticle excitations such as their dispersion and lifetime are significantly modified due to electron-electron interactions.\cite{TheBook}

Great attention has been paid in recent years to band-structure effects involving the spin degree of freedom.\cite{WinklerBook} The strength and form of spin-orbit interaction (SOI) can be controlled in two-dimensional liquids through the choice of materials, the type of carriers (electrons/holes), and details of the confinement potential. For example, it is possible to change the coupling constant with external gates.\cite{Datta1989,Nitta1997,Johnson2009} In addition to detailed studies of single-particle properties, the problem of understanding the effects of SOI in the presence of Coulomb interactions is a topic of ongoing investigations.

Quasiparticle properties in the presence of SOI\cite{Chen1999,Saraga2005,Nechaev2009,Nechaev2010,Agarwal2011} have been examined primarily accounting for Rashba\cite{Bychov1984a,Bychov1984b} and/or Dresselhaus SOI,\cite{Dresselhaus1955,Dyakonov1986} which are dominant in electronic systems. The SOI results in two distinct spin subbands, with two associated Fermi surfaces. The effects on quasiparticles are usually very small; at each of the two Fermi surfaces the quasiparticle dispersion\cite{Chen1999,Saraga2005,Agarwal2011} and lifetime\cite{Saraga2005,Nechaev2009,Nechaev2010} are almost unaffected by SOI, except in the case of very large SOI coupling.\cite{Nechaev2010,Berman2010} In fact, it was found with Rashba SOI that the corrections to these quantities linear in the SOI coupling are absent.\cite{Saraga2005} Although explicit calculations are performed within perturbative approximation schemes, notably the random-phase approximation (RPA),\cite{Saraga2005,Nechaev2009,Nechaev2010} the SOI leading-order cancellation is valid non-perturbatively (to all orders in $r_s$).\cite{ChesiExact} Similar arguments hold for other physical quantities.\cite{ChesiExact,Abedinpour2010} For example, values of the ground-state energy obtained with Monte Carlo simulations\cite{Ambrosetti2009} for up to $r_s =20$ could be reproduced with excellent accuracy by simply neglecting SOI corrections to the exchange-correlation energy.\cite{ChesiHighDens} Noticeable exceptions exhibiting larger SOI effects are spin-textured broken symmetry phases,\cite{ChesiHFstates,Juri2008} non-analytic corrections to the spin susceptibility,\cite{Zak2010,Zak2011} and the plasmon dispersion.\cite{Agarwal2011} All these examples involve the presence of spin polarization (either directly\cite{ChesiHFstates,Juri2008,Zak2010} or indirectly\cite{Agarwal2011}), in which case the arguments of Ref.~\onlinecite{ChesiExact} do not apply.

Another interesting situation occurs when the SOI has a nonlinear dependence on momentum, thus cannot be written as a spin-dependent gauge potential.\cite{Aleiner2001,Coish2006,Abedinpour2010} Then, the approximate cancellations mentioned above are not expected. Winkler has shown that the dominant SOI induced by heterostructure asymmetry is cubic in momentum for heavy holes in III-V semiconductors.\cite{Winkler2000} This theoretical analysis was later shown to be in good agreement with magnetoresistance experiments.\cite{Winkler2002} Recently, the relevance of this cubic-in-momentum model was supported by the anomalous sign and magnetic-field dependence of spin polarization in quantum point contacts.\cite{Rokhinson2004,ChesiQPC} SOI quadratic in momentum can also be induced for heavy holes by an in-plane magnetic field.\cite{Bulaev2007,Chesi07April}

That these band-structure effects can substantially modify standard many-body results is confirmed by recent Shubnikov-de Haas oscillation measurements in low-density hole systems.\cite{Winkler2005,Chiu2011} For example, a surprisingly small Coulomb enhancement of the $g$-factor has been reported \cite{Winkler2005} and puzzling results have also been obtained for the effective masses $m_\pm$ of the two hole-spin sub-bands ($\sigma=\pm$).\cite{Chiu2011} A mechanism for the small $g$-factor enhancement was suggested in Ref.~\onlinecite{Chesi07April}: if SOI strongly distorts the ground state spin structure, the exchange energy becomes ineffective in promoting full polarization of the hole system. 

\begin{figure}
  \centering
 \includegraphics[width = 0.45\textwidth]{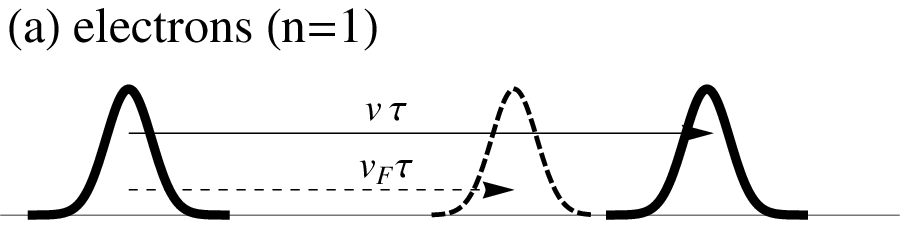}\\
 \vspace{0.2cm}
 \includegraphics[width = 0.45\textwidth]{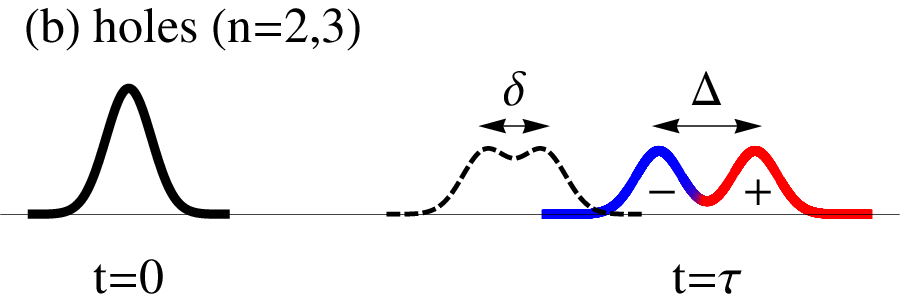}
   \caption{\label{fig:ElectronsVsHoles} (Color online) Motion of a wavepacket for a fixed time $\tau$. In (a) the SOI is not seen, because the difference in the velocities $v_\pm$ of the two spin components is too small. Panel (a) applies without SOI or to electrons with Rashba SOI ($n=1$). In (b) we show the effect of the non-linear SOI present in hole systems ($n=2,3$). Since the two spin branches have significantly different velocities $v_{\pm}$, there is an appreciable separation $\Delta=|v_{+}-v_{-}|\tau$. We also illustrate the propagation of non-interacting wavepackets (dashed) for the same time $\tau$. The effect of interactions (at high density) is to enhance both the average velocity ($v > v_F$) and the separation of spin-components: $\Delta > \delta=|v^0_{+}-v^0_{-}|\tau$. }
\end{figure}

In this paper, we focus on the effective masses $m_\pm$. We note that $m_\pm$ are directly related to the quasiparticle group velocities $v_\pm$ at the Fermi surfaces. With this in mind, we find it more transparent to discuss the effects of SOI and electron-electron interactions on $m_\pm$ in terms of wavepacket motion, as illustrated in Fig.~\ref{fig:ElectronsVsHoles}.  If an unpolarized wavepacket is injected at the Fermi energy (with average momentum along a given direction), the subsequent motion is very different depending on whether the SOI is linear ($n=1$) or non-linear ($n=2,3$). In the former case ($n=1$), corresponding to electrons, we have $v_+\simeq v_-$ and the motion is essentially equivalent to the case without SOI. In contrast, for holes with strong SOI, $v_+ \neq v_-$ and the two spin components become spatially separated. The separation between the two spin components can become quite sizable, and it should be possible to observe such an effect with, e.g., Faraday-rotation imaging techniques.\cite{Crooker1996,Kikkawa1997} Electron-electron interactions, in addition to modifying the average velocity $v=(v_++v_-)/2$ (an effect which is well-known without SOI\cite{TheBook, Tan2005,Drummond2009}), are also reflected on the velocity difference $(v_+-v_-)$ between the two spin branches. 

%=====================================================================================Cartoon
%==============================================================================================

With these motivations in mind, we pursue a study of the quasiparticle group velocity in the presence of linear Rashba and non-linear SOI. An accurate analytical treatment of the electron-electron interactions can be carried out at high density, and we restrict ourselves to this interesting limit. The regime of strong electron-electron interactions is much more difficult to treat (see Ref.~\onlinecite{Drummond2009} for a Monte Carlo study without SOI) but it represents a relevant topic for future investigations ($r_s \sim 6 - 12$ in Refs.~\onlinecite{Winkler2005,Chiu2011}). 

This paper is organized as follows: In Sec.~\ref{sec:Noninteracting} we introduce a model Hamiltonian including a generalized SOI\cite{Chesi07April} and we briefly review its non-interacting properties, demonstrating that injected holes will separate into spin-polarized wavepackets. The Coulomb interaction is treated in Sec.~\ref{sec:screenedHF} by extending the classic treatment of Ref.~\onlinecite{Janak1969}. We describe several results of this screened Hartree-Fock approach in detail, focusing on the quasiparticle group velocities and the interplay of spin-orbit and Coulomb interaction effects. A discussion of these results is given in Sec.~\ref{sec:conclusion}. In Appendix~\ref{app:gneral_arg}, we give a general argument showing that corrections to the velocity from any linear-in-momentum SOI can always be neglected to lowest order. Finally, a number of technical details are provided in Appendices~\ref{app:occupation} and \ref{app:expansions}.

%=======================================================================================
\section{Noninteracting problem}\label{sec:Noninteracting}
%=======================================================================================

In the high-density limit, $r_s$ is small and the system is well-described by a non-interacting single-particle Hamiltonian. We consider here the following model,\cite{Chesi07April} including a generalized SOI with a linear-, quadratic-, and cubic-in-momentum dependence for $n=1,2,3$:
\begin{equation}\label{eq:H0}
H_{0} = \frac{p^2}{2m} + i\gamma \frac{p_{-}^n\sigma_{+} -p_{+}^n\sigma_{-} }{2},
\end{equation}
where ${\bf p}$ is the momentum operator, $m$ the band mass, $p_{\pm} = p_{x} \pm i p_{y}$, $\sigma_{\pm} = \sigma_{x} \pm i \sigma_{y}$, and $\gamma$ the generalized spin-orbit coupling, with $\boldsymbol{\sigma}$ the vector of Pauli matrices. The physical justification of this model has been given in Ref.~\onlinecite{Chesi07April}: the $n=1$ Hamiltonian contains the Rashba SOI present in electronic systems,\cite{Bychov1984a,Bychov1984b} while $n=3$ includes the analogous term generated by an asymmetric confinement potential for holes.\cite{Winkler2000,Winkler2002,ChesiQPC} Finally, the $n=2$ case is also relevant for holes, in the presence of an in-plane magnetic field.\cite{Chesi07April,Bulaev2007} While SOI terms of different form generally coexist (see Appendix~\ref{app:gneral_arg}), we assume here that one $n$ value is dominant. This greatly simplifies the problem by preserving the isotropy of the electron liquid in the x-y plane and is a good approximation for several relevant situations. For example, it was found for holes\cite{Winkler2000,Winkler2002,ChesiQPC} that the $n=3$ term can be much larger than corrections to the SOI due to bulk inversion asymmetry.\cite{WinklerBook,Rashba1988} Diagonalizing $H_0$ yields the energy spectrum
\begin{equation}\label{eq:Ek-non-interacting}
E_{\sigma}^{0} (k) = \frac{\hbar^2 k^2}{2m}+ \sigma \gamma \hbar^n k^n,
\end{equation}
with $\sigma = \pm$ labeling the two chiral spin branches. The corresponding eigenfunctions are\cite{Chesi07April}
\begin{equation}\label{eq:spinors}
\varphi_{ \mathbf{k} \sigma}(\mathbf{r}) = \frac{ e^{i \mathbf{k} \cdot \mathbf{r} }}{\sqrt{2 L^2}} \begin{pmatrix} 1 \\ i \sigma e^{in \theta_{\mathbf{k}}} \end{pmatrix},
\end{equation}
where $\mathbf{k}$ is a wavevector in the x-y plane, $\theta_{\bf k}$ is the angle $\mathbf{k}$ makes with the x-axis, and $L$ is the linear size of the system. 

It is useful at this point to introduce a dimensionless quantity $g$ characterizing the strength of the spin-orbit coupling:\cite{Chesi07April}
\begin{equation}\label{eq:g_def}
g=\frac{\gamma \hbar^n k_F^n}{E_F},
\end{equation}
where $E_F=\hbar^2 k_F^2/2m$ is the Fermi energy without SOI, written in terms of the Fermi wavevector $k_F=\sqrt{2\pi n_s}=\sqrt{2}/(a_\mathrm{B}r_s)$.  While $\gamma$ has different physical dimensions for each value, $n=1,2,3$, the dimensionless coupling $g$ always gives the ratio of the spin-orbit energy to the kinetic energy. The coupling $g$ thus plays a role analogous to that of $r_s$ for the Coulomb interaction. Taking $\gamma$ to be independent of the density (this is not always the case\cite{WinklerBook}), then from Eq. \eqref{eq:g_def} we have $g\propto k_F^{n-2}\propto r_s^{2-n}$ (since $E_F\propto k_F^2$).  This suggests that in the high-density limit ($r_s\to 0$), the effects of SOI are suppressed for electrons ($n=1$), but remain constant ($n=2$) or are enhanced ($n=3$) for holes.  This simple estimate already indicates a qualitative difference for holes relative to electrons.  We will see that this difference is indeed significant in the following sections.

In the presence of SOI the two spin bands $(\sigma = \pm)$ have different densities $n_{\pm}$, giving the total 2D sheet density $n_s= n_{+}+n_{-}$. Keeping $n_s$ (thus $k_F$) fixed gives a constraint on the Fermi wavevectors $k_{\pm}$ for the $\sigma=\pm$ bands, $k_{+}^2+k_{-}^2 = 2 k_{F}^2$. We can then characterize the solution to this equation with a single parameter $\chi$:
\begin{equation}
k_{\pm} = k_{F} \sqrt{1 \mp \chi}.
\end{equation}
The parameter $\chi = (n_{-}-n_{+})/n_s$ gives the chirality and is determined by both the SOI and electron-electron interaction.\cite{comment_gen_chi} We will assume for definiteness that $\gamma \geq 0$ such that $\chi \geq 0$ and $k_+ \leq k_-$. For a fixed generalized spin-orbit coupling $\gamma$, the non-interacting value of $\chi$ can be determined by setting the Fermi energies of the two bands equal, i.e., $E^0_{+}(k_+)=E^0_-(k_-)$. This equation immediately gives a relationship between $g$ and $\chi$,
\begin{equation}\label{eq:Gamma-n}
g = \frac{2\chi}{(1+\chi)^{n/2} + (1-\chi)^{n/2}}.
\end{equation}
We denote the solution of Eq.~(\ref{eq:Gamma-n}) by $\chi_0 (g)$, which gives the non-interacting Fermi wavevectors $k_\pm^0 = k_F \sqrt{1\mp \chi_0}$. Explicit expressions for $\chi_0(g)$ are given in Ref.~\onlinecite{Chesi07April}. We only cite here the small-$g$ behavior, which is easily found from Eq.~(\ref{eq:Gamma-n}):
\begin{equation}\label{eq:chi0_low_order}
\chi_0(g) \simeq g.
\end{equation}

We are mainly interested here in the properties of the quasiparticles and, in particular, their group velocities $v_{\pm}$. The fact that the dispersion relation \eqref{eq:Ek-non-interacting} is a function only of the magnitude $k$ is a consequence of the model being isotropic with respect to $\mathbf{k}$, which allows us to discuss the magnitude of the group velocity on the Fermi surfaces,
\begin{equation}\label{eq:vF0}
v_{\pm}^{0} = \frac{1}{\hbar} \frac{ \partial E_{\pm}^{0}(k)}{\partial k} \Big |_{k=k^0_{\pm}} = \frac{\hbar k^0_{\pm}}{m} \pm n \gamma (\hbar k^0_{\pm})^{n-1}.
\end{equation}
The above expression can be evaluated explicitly in terms of the non-interacting chirality $\chi_0(g)$:
\begin{eqnarray}\label{vF_nonint}
\frac{v_{\pm}^{0}}{v_{F}}= && \sqrt{1\mp \chi_0(g)} \pm \frac{n}{2} g [1\mp \chi_0(g)]^{(n-1)/2} \nonumber \\
&& \simeq 1\pm \frac{g}{2}(n-1),
\end{eqnarray}
where $v_F=\hbar k_F/m$ is the Fermi velocity in the absence of SOI and in the second line we have expanded the result to lowest order in $g$, by making use of Eq.~(\ref{eq:chi0_low_order}). Equation~(\ref{vF_nonint}) shows that there is no relative difference in group velocity for the $\sigma = \pm$ bands when $n=1$. In contrast, when $n=2,3$ a sizable correction linear in $g$ is present.  This difference reflects itself on the evolution of an initially unpolarized wavepacket injected at the Fermi surface, schematically illustrated in Fig.~\ref{fig:ElectronsVsHoles}. While for electrons the wavepacket remains unpolarized, for holes the two spin components spatially separate with time. Electron-electron interactions modify the non-interacting velocities $v^0_\pm$, but the qualitative difference introduced by SOI between electrons ($n=1$) and holes ($n=2,3$) remains essentially unchanged.  A similar behavior holds for other properties of the electron liquid as well:\cite{ChesiExact} vanishing corrections to lowest order in $g$ were found for the quasiparticle lifetime,\cite{Saraga2005,Nechaev2010} the occupation,\cite{Chesi07April} and the exchange-correlation energy,\cite{ChesiHighDens} if only Rashba SOI ($n=1$) is included. 

Finally, another relevant quasiparticle observable is the effective mass.  This has recently been measured in hole systems through experiments on quantum oscillations.\cite{Chiu2011} This physical quantity is simply given by $m_{\pm}  = \hbar k_{\pm}/ v_{\pm}$ and is thus essentially equivalent to $v_\pm$. 

%=======================================================================================
\section{Screened Hartree-Fock approximation}\label{sec:screenedHF}
%=======================================================================================

Realistically, charged particles interact through the Coulomb potential so it is interesting to understand how the presence of SOI modifies the behavior of the quasiparticles. The fully interacting Hamiltonian is given by:
\begin{equation}\label{eq:H}
H=\sum_i H_0^{(i)} +\frac12 \sum_{i\neq j} \frac{e^2}{\epsilon_r |{\bf r}_i -{\bf r}_j|},
\end{equation}
where $H_0^{(i)}$ (for electron $i$) is as in Eq.~(\ref{eq:H0}) and the presence in (\ref{eq:H}) of a uniform neutralizing background is understood. Although many sophisticated techniques exist to approach this problem,\cite{TheBook,Drummond2009} the simplest approximation to the quasiparticle self-energy is obtained by only including the exchange contribution:
\begin{equation}
E_{\sigma}(k)  = E_\sigma^{0}(k) + \Sigma^x_{ \sigma}(k), \label{eq:dispersion}
\end{equation}
where
\begin{equation} \label{eq:exchange}
\Sigma^{x}_{\sigma}(k) =  - \sum_{\mathbf{k}' \sigma'} \frac{(1+\sigma \sigma' \cos{n \theta'})}{2 L^2} n_{\mathbf{k}' \sigma'} V(|\mathbf{k}-\mathbf{k}'|)  .
\end{equation}
Here, $n_{\mathbf{k'} \pm} = \Theta(k_{\pm}-k')$ is the occupation at $T=0$ for the $\sigma=\pm$ band, respectively, with $\Theta(x)$ the Heaviside step function. The first factor in the summation of Eq.~(\ref{eq:exchange}), involving the angle $\theta'$ between $\mathbf{k'}$ and $\mathbf{k}$, arises from the scalar product of the non-interacting spinors [Eq.~(\ref{eq:spinors})], and takes into account the specific nature of the spin-orbit interaction ($n=1,2,3$). 

To lowest order, $V(q)$ is the Fourier transform of the bare Coulomb potential,  $2\pi e^2/(\epsilon_r q)$. As is well known,\cite{TheBook} this form of the Coulomb interaction leads to an unphysical divergence in the quasiparticle velocity. By considering an infinite resummation in perturbation theory, the screening of the Coulomb interaction removes the divergence, e.g., in the RPA approximation. Finally, by approximating the dielectric function in the effective interaction by its zero-frequency long-wavelength limit, the RPA self-energy gives Eq.~(\ref{eq:exchange}) with
\begin{equation}
V(q) = \frac{2 \pi e^2/ \epsilon_r}{q + \sqrt{2} r_{s} k_{F}}. \label{eq:coulomb}
\end{equation}
This screened Hartree-Fock approximation with SOI,\cite{Chen1999,Agarwal2011} notwithstanding its simplicity, becomes accurate in the high-density limit.

%=======================================================================================
\subsection{Renormalized occupation}\label{sec:occupation}
%=======================================================================================

By including the SOI, we can verify that Eq.~(\ref{eq:exchange}) gives the correct high-density behavior for the Fermi wavevectors $k_\pm$. These are modified by electron-electron interactions from their non-interacting values $k_\pm^0$.\cite{ Schliemann2006, Chesi07April} From Eq.~(\ref{eq:exchange}), $k_\pm$ can be obtained by equating the chemical potentials in the two spin branches:
\begin{equation}
E_+(k_+) = E_- (k_-).
\end{equation}
After taking the continuum limit, this equation is rewritten in dimensionless form as follows:
\begin{eqnarray} \label{eq:Gamma-n-app}
(y_{+}^n + y_{-}^n) g =  && 2\chi   +  \frac{r_s}{\sqrt{2}}
 \sum_{\sigma \sigma'}  \int_0^{2\pi} \! \frac{d\theta}{2\pi} \! \int_{0}^{y_{\sigma'}} \! dy  \nonumber\\
&& \times \frac{y(\sigma+\sigma' \cos{n\theta})}{\sqrt{y^2+y_{\sigma}^{2} - 2y y_{\sigma} \cos{\theta}}+\sqrt{2}r_s} ,~~
\end{eqnarray}
where we have rescaled the wavevectors $k=k_F y$ and defined $y_\pm =\sqrt{1\mp \chi}$. The integral on the right-hand side is the correction from the exchange term and we have verified that Eq.~(\ref{eq:Gamma-n-app}) gives Eq.~(\ref{eq:Gamma-n}) for $r_s=0$. We note that, for a given value of the SOI, $\chi(r_s,g)$ enters in a rather complicated way in Eq.~(\ref{eq:Gamma-n-app}), being involved in the integration limits of the exchange term as well as the integrand. In practice, instead of solving Eq.~(\ref{eq:Gamma-n-app}) for $\chi$, it is convenient to evaluate $g$ for a given value of $\chi$ and numerically invert the function $g(r_s,\chi)$.

The values of $k_\pm = k_F y_\pm$ were obtained in Ref.~\onlinecite{Chesi07April} through a different procedure, i.e., by minimizing the total energy (including the exchange contribution) of non-interacting states.\cite{comment_variational} Although both methods are unreliable at $r_s>1$, they both become accurate in the high-density limit, $r_s<1$.  In fact, the only difference in the two approaches is due to the presence of the Thomas-Fermi screening wavevector in Eq.~(\ref{eq:Gamma-n-app}) and, by neglecting $\sqrt{2}r_s$ in the denominator of the second line, the equation from the variational treatment is recovered. In particular, expanding Eq.~(\ref{eq:Gamma-n-app}) at small $r_s$ and $g$ gives the same result found in Ref.~\onlinecite{Chesi07April}:
\begin{equation}\label{chi_interacting}
\chi(r_s,g) \simeq g\left( 1-\frac{\sqrt{2}r_s}{\pi}\sum_{j=0}^n \frac{1}{2j-1}\right).
\end{equation}
Details of the derivation of Eq. \eqref{chi_interacting} are given in Appendix~\ref{app:occupation}. Two salient features of Eq.~(\ref{chi_interacting}) are:\cite{Chesi07April} (i) For $n=1$ there is no correction to the noninteracting result $\chi(r_s,g) \simeq g$. On the other hand, the linear dependence on $g$ is actually modified by electron-electron interactions at $n=2,3$. (ii) The effect of the electron-electron interactions is a reduction of $\chi(r_s,g)$ from the non-interacting value. This result could be rather surprising, having in mind the well-known enhancement of spin polarization caused by the exchange energy\cite{TheBook} (when the spin-splitting is generated by a magnetic field). However, $\chi$ does not correspond here to a real spin-polarization, which is zero.  Instead, $\chi$ is simply related to the population difference of the two chiral spin subbands.

%=======================================================================================
\subsection{Quasiparticle velocity\label{sec:v_general}}
%=======================================================================================

In the screened Hartree-Fock approximation, the group velocities at the Fermi surfaces are given by:
\begin{eqnarray}\label{eq:groupvelocity}
v_{\pm} = && \, \frac{1}{\hbar} \frac{ \partial E_{k \pm}}{\partial k} \Big |_{k=k_{\pm}}\nonumber  \\
 = && \, \frac{\hbar k_{\pm}}{m} \pm n \gamma (\hbar k_{\pm})^{n-1} \nonumber \\
&& -\sum_{\mathbf{k}' \sigma'} \frac{(1\pm \sigma' \cos{n \theta'})}{2L^2} n_{\mathbf{k}' \sigma'} \left[\frac{\partial }{\partial k}V(|\mathbf{k}\!-\!\mathbf{k'}|)\right]_{k=k_\pm}. 
\end{eqnarray}
In general, we can discuss all corrections to $v_\pm$ by introducing the following notation:
\begin{equation}\label{eq:corrections_def}
\frac{v_{\pm}}{v_F} =1+ \delta v(r_s)+ \delta v_{\pm}^{0}(g) + \delta v_{\pm}(r_s,g),
\end{equation}
where $\delta v(r_s)$ is the (spin-independent) correction due to electron-electron interactions at $g=0$, which has been the subject of many theoretical and experimental studies (see, e.g., Ref.~\onlinecite{TheBook}, \onlinecite{Chiu2011,Janak1969,Tan2005,Drummond2009}, and references therein). In the approximation (\ref{eq:groupvelocity}), it is given by\cite{Janak1969}
\begin{equation}\label{eq:v_janak}
\delta v(r_s)= -\frac{\sqrt{2}r_s}{\pi}+\frac{r_s^2}{2}+\frac{r_s(1-r_s^2)}{\sqrt{2}\pi}\frac{\cosh^{-1}(\sqrt{2}/r_s)}{\sqrt{1-r_s^2/2}}.
\end{equation}
As is known,\cite{TheBook} this approximation gives the correct leading behavior at small $r_s$: $\delta v \simeq - (r_s \ln r_s)/(\sqrt{2}\pi)$. The second nontrivial term in Eq. \eqref{eq:corrections_def} is the non-interacting correction purely due to SOI
\begin{equation}
\delta v^0_\pm(g)=\frac{v^0_{\pm}(g)}{v_F}-1,
\end{equation}
which only depends on $g$ [see Eq.~(\ref{vF_nonint})]. Finally, $\delta v_{\pm}(r_s,g)$ collects all remaining corrections. 

A pictorial representation of the physical meaning of the three terms is shown in Fig.~\ref{fig:ElectronsVsHoles} for an unpolarized wavepacket injected at the Fermi energy. At high density, as seen in Eq.~(\ref{eq:v_janak}) and illustrated in Fig. \ref{fig:ElectronsVsHoles}(a), the group velocity is larger than without electron-electron interactions. In Fig. \ref{fig:ElectronsVsHoles}(b) we depict the generic situation with SOI.  In the presence of SOI, the two spin branches have different group velocities.  An initially unpolarized wavepacket then splits into its two spin components. Both the SOI and electron-electron interactions influence the relative velocity $(v_{+} - v_{-})$.  The separation after a time $\tau$ is $\delta = |v^0_+ -v_-^0| \tau $ for the non-interacting case and is modified by $\delta v_\pm$ with electron-electron interactions: $\Delta = |v_+ -v_-|  \tau $ [see Fig.~\ref{fig:ElectronsVsHoles}(b)]. 

The presence of `interference' terms, $\delta v_\pm(r_s,g)$, becomes clear from Eq.~(\ref{eq:groupvelocity}). These terms are due to the interplay of many-body interactions with SOI.  A first contribution to $\delta v_{\pm}(r_s,g)$, which we refer to as the `self-energy contribution', comes directly from the exchange integral [third line of Eq.~(\ref{eq:groupvelocity})]:  due to the presence of two distinct Fermi wavevectors $k_\pm$, the result obviously contains corrections to Eq.~(\ref{eq:v_janak}) which depend on $g$ (in addition to $r_s$). A second contribution to $\delta v_\pm(r_s,g)$ comes indirectly from the non-interacting part and we refer to it as the `repopulation contribution'.  Since the Fermi wavevectors $k_\pm$ are modified from the non-interacting values $k_\pm^0$, the second line of Eq.~(\ref{eq:groupvelocity}) gives a result distinct from $v^0_{\pm}$. The sign of this repopulation contribution is easily found by noting that, as discussed in Sec.~\ref{sec:occupation}, the exchange energy reduces the value of $\chi$ (for $n=2,3$). This corresponds to an increase (decrease) of $k_+$ ($k_-$), i.e., a positive (negative) correction to $v_{+}$ ($v_{-}$). The effect would be to enhance the difference in velocity between the two branches, as illustrated in Fig.~\ref{fig:ElectronsVsHoles}(b) ($\Delta>\delta$). However, to establish the ultimate form and sign of $\delta v_\pm(r_s,g)$ requires a detailed calculation of both self-energy and repopulation contributions, which is presented below for some interesting cases. The total spin-dependent part of the velocity, $\delta v_\pm^0 + \delta v_\pm$, can then be compared to the simple non-interacting effect, $\delta v_\pm^0$.

%=======================================================================================
\subsection{$n=1$: higher-order corrections in $g$ \label{sec:velocity_n1}}
%=======================================================================================

We begin the analysis of Eq.~(\ref{eq:groupvelocity}) by rewriting it in a more explicit way. To this end, we use 
\begin{equation}
\frac{\partial}{\partial k} V(|\mathbf{k}-\!\mathbf{k}'|)  = -\hat{\mathbf{k}} \cdot \frac{\partial}{\partial{\mathbf{k}'}} V(|\mathbf{k}-\mathbf{k}'|),
\end{equation}
where $\hat{\bf k}={\bf k}/k$. This allows us to integrate Eq.~\eqref{eq:groupvelocity} by parts, which leads to:
\begin{eqnarray}\label{eq:groupvelocity_explicit}
&& \frac{v_{\pm}}{v_{F}} =   y_\pm \pm \frac{g}{2} n y_\pm ^{n-1} \nonumber \\
&& +   r_s \sum_{\sigma} \int_0^{2\pi} \frac{d \theta'}{8\pi} \frac{\sqrt{2} \cos{\theta'}(1\pm \sigma \cos{n \theta'})y_\sigma }{\sqrt{y_\pm^2+y_\sigma^2 -2 y_\pm y_\sigma \cos{\theta'}} + r_{s}\sqrt{2}} \nonumber\\
&& \pm r_s \int_0^{2\pi} \frac{d \theta'}{8\pi} \int_{y_{+}}^{y_{-}}\frac{\sqrt{2} n\sin{n\theta'}\sin{\theta'} \,  dy' }{\sqrt{y_{\pm}^2 +{y'}^2 - 2 y' y_{\pm} \cos{\theta'}} + r_s \sqrt{2}} \nonumber \\
&& = L_1 + L_2 + L_3.
\end{eqnarray} 
Notice that, in the integration by parts of Eq.~(\ref{eq:groupvelocity}) in the continuum limit, two types of terms enter: those corresponding to the second line of Eq.~(\ref{eq:groupvelocity_explicit}) ($L_2$) involve the derivative of $n_{\mathbf{k'}\pm}$. This results in a delta function in the $dk'$ integral, which can then be easily evaluated. Thus, only the integral in $d\theta'$ is left. The second type of term involves the derivative of $\cos{n \theta'}$:
\begin{equation}
\hat{\mathbf{k}} \cdot \frac{\partial}{\partial \mathbf{k'}} \cos{n \theta'} = \frac{n}{k'} \sin{n \theta'} \sin{\theta'},
\end{equation}
and indeed this angular factor appears in the third line of Eq.~(\ref{eq:groupvelocity_explicit}) ($L_3$). 

Eq.~(\ref{eq:groupvelocity_explicit}) can always be evaluated numerically, after obtaining the values of $\chi$ (thus $y_\pm=\sqrt{1\mp \chi}$) from Eq.~(\ref{eq:Gamma-n-app}). By specializing to the small $g,r_s$ limit for $n=1$ SOI, it is known that the linear-in-$g$ correction to $v_{\pm}$ vanishes.\cite{Saraga2005,ChesiExact} Thus, an expansion to second order in $g$ has to be performed, which has been done in Ref.~\onlinecite{Saraga2005} in the context of the RPA treatment of the quasiparticle properties. To verify the validity of the simpler screened Hartree-Fock procedure, it is interesting to perform the same expansion for our Eq.~(\ref{eq:groupvelocity_explicit}).

In fact, the two calculations bear some similarities since $L_2$ is the same as the boundary term $B_{\rm boundary}^{(u\to 0_+)}$ of the RPA treatment, see Eq.~(69) of Ref.~\onlinecite{Saraga2005}. Thus, we can borrow the expansion in small $g,r_s$:
\begin{equation}\label{eq:L2}
L_2 \simeq -\frac{r_s}{\sqrt{2}\pi} \left( \ln \frac{r_s}{2\sqrt{2}}+2 \pm \frac23 \, g -\frac{g^2}{8} \ln g \right);\quad (n=1),
\end{equation}
where terms of order $O(r_s^2,r_s g^2)$ have been omitted. To the same order of approximation, we have for $L_1$:
\begin{equation}\label{eq:L1}
L_1 \simeq  1-\frac{g^2}{8};\quad (n=1),
\end{equation}
which can also be obtained by expanding Eq.~(\ref{vF_nonint}) (with $n=1$). Since $\chi$ only receives $O(r_s g^3 \ln g)$ corrections from electron-electron interactions,\cite{ChesiHighDens} it is sufficient to use the non-interacting value, $\chi_0(g)$, to this order of approximation. Thus, the repopulation induced by electron-electron interactions has a negligible effect on $v_{\pm}$ in this case. The situation will be different for $n=2,3$. Expanding $L_3$ for small $r_s,g$ yields
\begin{equation}\label{eq:L3}
L_3 \simeq \pm \frac{\sqrt{2}r_s}{3\pi}g;\quad (n=1),
\end{equation}
which cancels the linear-in-$g$ term of Eq.~(\ref{eq:L2}), as expected for $n=1$. 

We note that in the screened Hartree-Fock approximation we are able to obtain an analytic result for the leading term of $L_3$, and some details of the derivation can be found in Appendix~\ref{app:expansions}. In contrast, in the RPA treatment of Ref.~\onlinecite{Saraga2005} it was not possible to expand the more involved corresponding term, $B_{\rm int}$, in a fully analytic fashion. The cancellation of the linear-in-$g$ contribution was indicated by a general argument\cite{Saraga2005,ChesiExact} and confirmed through numerical study. Additionally, the absence of higher-order terms which modify Eq.~(\ref{eq:L2}) was inferred numerically in Ref.~\onlinecite{Saraga2005}. Although the final results of both approaches (RPA and screened Hartree-Fock) agree, the situation is clearly more satisfactory within the framework of Eq.~(\ref{eq:groupvelocity_explicit}), since cancellation of the linear term in Eq.~(\ref{eq:L2}) can be checked exactly and the expansion can be carried out to higher order systematically. The final result, computed to higher order in $g$, reads (for $n=1$):
\begin{eqnarray}\label{eq:n1_v_final}
 \frac{v_{\pm}}{v_F} && \simeq 1+  \delta v(r_s)+\delta v^0_\pm(g)  \nonumber \\
&& + \frac{\sqrt{2} r_s}{16\pi}  \left[ g^2 \left(\ln \frac{g}{8} + \frac32\right) \pm  \frac{g^3}{6}\left(\ln\frac{g^7}{8^5}+ \frac{319}{20}\right)\right].~~~~~~~
\end{eqnarray}
%
%================================================difference in group velocity.
 \begin{figure}
  \centering
   \includegraphics[width = 0.45\textwidth]{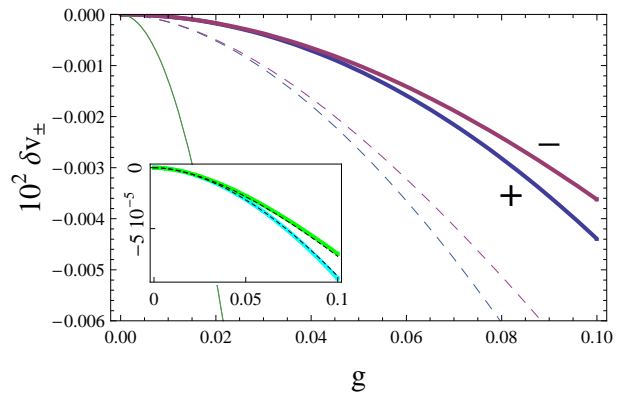}
   \caption{\label{fig:n=1check} (Color online) Thick solid lines: the corrections $\delta v_\pm (r_s,g)$ evaluated by numerical integration of Eq. \eqref{eq:groupvelocity_explicit}. Dashed lines: approximation to $\delta v_\pm (r_s,g)$, given by the second line of of Eq.~(\ref{eq:n1_v_final}). In the main plot, we take $r_s=0.1$, for which Eq.~(\ref{eq:n1_v_final}) is not very accurate. Very good agreement is obtained at small values of $r_s$ (inset, with $r_s=0.001$). The thinner solid line in the main plot is the total correction due to spin-orbit coupling: $\delta v_\pm^0(g)+\delta v_\pm (r_s,g)$. It is dominated by the non-interacting effect and the spin-splitting is not visible. The value of $\delta v_\pm^0(g)$ at $g=0.1$, outside the plot range, is $\sim 1 \%$ (all corrections are in units of $v_F$).}
\end{figure}
%=======================================================================================
%
The second line of Eq.~(\ref{eq:n1_v_final}) represents the expansion of $\delta v_\pm (r_s,g)$ including all terms up to $O(r_s g^3$).  For $n=1$, the non-interacting result gives the same velocity for both spin branches [$\delta v^0_\pm(g)$ is actually independent of $\pm$ for this particular model].  The $O(r_s g^3)$ term in Eq.~(\ref{eq:n1_v_final}) is therefore quite interesting. It shows that a small difference in velocity exists.  This is a genuine effect of electron-electron interactions. 

The accuracy of Eq.~(\ref{eq:n1_v_final}) can be seen in Fig.~\ref{fig:n=1check}: it becomes accurate at very small values of $r_s$ (as shown in the inset) while at larger more realistic values of $r_s$ it still gives the correct magnitude of the effect. The numerical example of Fig.~\ref{fig:n=1check} also shows that $\delta v_\pm (r_s,g)$ is very small. In fact, it is generally much smaller than the non-interacting correction: since $\delta v^0_{\pm}(g)\simeq -g^2/8$ [see Eq.~(\ref{eq:L1})] the $~r_s g^2 \ln g$ term becomes larger only if $r_s \ln g \gg 1$. This condition is only satisfied at extremely small values of $g$ (if $r_s$ is small as well) at which SOI effects are hardly of any relevance. Thus, Eq.~(\ref{eq:n1_v_final}) implies that the effects of SOI and electron-electron interactions are essentially decoupled for $n=1$.  This picture changes substantially for hole systems ($n=2,3$), as we show in the next section.

%=======================================================================================
\subsection{Corrections to the velocity for $n=2,3$ \label{sec:velocity_n23}}
%=======================================================================================

We now apply the discussion of the previous section to the SOI more appropriate for holes and point out some important differences.  Expansions for small $r_s$ and $g$ are given in Appendix \ref{app:expansions}. From Eqs.~(\ref{eq:Isum_2}) and (\ref{eq:Isum_3}) and using $\chi=g+O(r_s g, g^3)$ [Eq. \eqref{chi_interacting}], we find the following expressions for the self-energy contribution. For $n=2$: 
\begin{equation}\label{eq:L23_2}
L_2+L_3 \simeq 
\, \delta v(r_s)+\frac{\sqrt{2} r_s}{4\pi}  \left[ \pm \frac{g}{3} + g^2 \left(\ln\frac{g}{8}+2\right)\right],
\end{equation}
and for $n=3$: 
\begin{equation}\label{eq:L23_3}
L_2+L_3 \simeq 
\, \delta v(r_s)+\frac{\sqrt{2} r_s}{4\pi}  \left[ \pm \frac{8}{15} g + \frac{g^2}{4} \left(9\ln\frac{g}{8}+\frac{613}{30}\right)\right].
\end{equation}
At variance with the case of $n=1$, the linear-in-$g$ term does not vanish here. Thus, we find an appreciable correction to the velocity.  This correction has opposite sign in the two branches and is positive for the $+$ (higher-energy) branch.

A second contribution to the $g$-linear correction comes from $L_1$. Expanding $L_1$ in terms of $\chi$ and using Eq.~(\ref{chi_interacting}) gives, for $n=2$,
\begin{equation}\label{eq:L1n2}
L_1 \simeq  1+ \delta v^0_\pm(g) + \frac{\sqrt{2} r_s}{4\pi} \left(\pm \frac23 g + g^2 \right);\quad (n=2),
\end{equation}
and for $n=3$:
\begin{equation}\label{eq:L1n3}
L_1 \simeq  1+ \delta v^0_\pm(g)+ \frac{\sqrt{2} r_s}{4 \pi}\left(\pm \frac{16}{15} g + \frac{32}{15} g^2 \right);\quad (n=3).
\end{equation}
Thus, the repopulation contribution is present in this case and has the sign discussed at the end of Sec.~\ref{sec:v_general} (it is positive for the $+$ branch). 

As it turns out, the self-energy, repopulation, and non-interacting contributions to the velocity have the same sign. The three contributions therefore have a cooperative effect in enhancing the difference in velocity between the two spin branches. Of course, based on the high-density theory presented here, we cannot tell if this conclusion holds at all densities. We also note that the $g$-linear term of the self-energy correction [Eq.~(\ref{eq:L23_2}) or (\ref{eq:L23_3})] is always half of the corresponding repopulation correction [Eq.~(\ref{eq:L1n2}) or (\ref{eq:L1n3})]. Again, we have not investigated if this curious relation only occurs within this approximation scheme or if it is more general. 

\begin{figure}
  \centering
   \includegraphics[width = 0.45\textwidth]{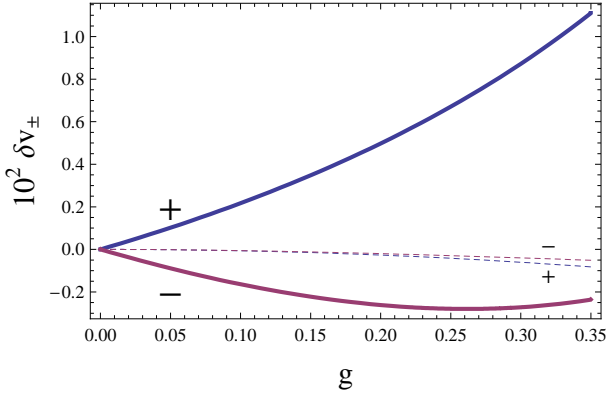}
   \caption{\label{fig:GroupVelocityDifference} (Color online) Plot of $\delta v_\pm (r_s,g)$ for $n=3$ (solid curves) and $n=1$ (dashed curves) as a function of the SOI strength $g$. We have taken $r_s=0.3$ in both cases. $\pm$ indicate the spin branch of each curve.}
\end{figure}

\begin{figure}
  \centering
   \includegraphics[width = 0.45\textwidth]{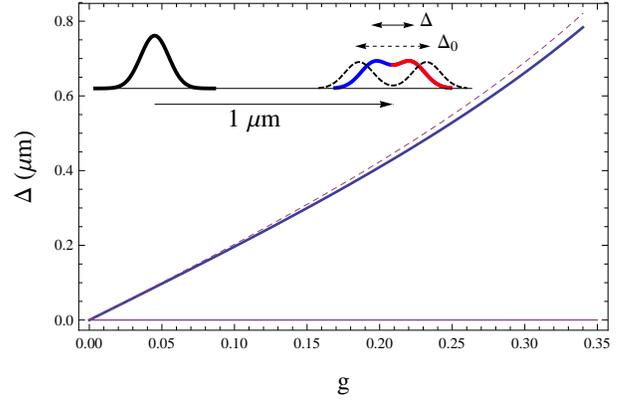}
   \caption{ \label{fig:SeperationFixedDistance} (Color online) Thick solid line: separation $\Delta$ between the two spin components of a hole wavepacket ($n=3$) at a fixed drift distance of $1~\mu {\rm m}$, see Eq.~(\ref{eq:Delta}). The dashed line is the non-interacting value $\Delta_0$. The thin solid line is $n=1$, indistinguishable from $\Delta=0$. We have assumed here $r_s=0.3$. The inset schematically illustrates the definitions of $\Delta$ and $\Delta_0$.}
\end{figure}

Finally, we give the complete result for $n=2$
\begin{eqnarray}\label{eq:n2_v_final}
 \frac{v_{\pm}}{v_F} && \simeq 1 +  \delta v(r_s)+\delta v^0_\pm(g)  \nonumber \\
&& +\frac{\sqrt{2} r_s}{4\pi}  \left[ \pm  g +  g^2 \left(\ln\frac{g}{8}+3\right)\right];\quad (n=2),~~~
\end{eqnarray}
and for $n=3$
\begin{eqnarray}\label{eq:n3_v_final}
 \frac{v_{\pm}}{v_F} && \simeq 1+  \delta v(r_s)+\delta v^0_\pm(g)  \nonumber \\
&& +\frac{\sqrt{2} r_s}{4\pi}  \left[ \pm \frac{8}{5} g + g^2 \left(9\ln\frac{g}{8}+\frac{869}{120}\right)\right];\quad (n=3),~~~
\end{eqnarray}
and show two numerical examples in Figs.~\ref{fig:GroupVelocityDifference} and \ref{fig:SeperationFixedDistance}. 
%
%================================================difference in group velocity.
%=======================================================================================
%
%===================================Seperation of wavepackets at fixed distance figure.
%=======================================================================================
%
Fig. \ref{fig:GroupVelocityDifference} is a comparison of $\delta v_\pm(r_s,g)$ for $n=1$ and $n=3$: it is clear that the dependence on $g$ is very weak for the electron case ($n=1$) and the magnitude of $\delta v_\pm(r_s,g)$ is much larger for holes ($n=3$) with SOI of comparable strength. Fig. \ref{fig:SeperationFixedDistance} shows the separation $\Delta$ between the two spin components of an initially unpolarized wave packet for a fixed travel distance of $1~\mu$m:
\begin{equation}\label{eq:Delta}
\Delta =2\,\frac{|v_{+}-v_{-}|}{v_{+}+v_{-}} \times (1~\mu{\rm m}).
\end{equation}
In addition to including both $n=1$ and $n=3$, in Fig.~\ref{fig:SeperationFixedDistance} we plot the non-interacting value $\Delta_0$, obtained by substituting $v_{\pm} \to v^0_{\pm}$ in Eq.~(\ref{eq:Delta}). For $n=1$ the non-interacting velocities $v_{\pm}^0$ are the same ($\Delta_0=0$). The effect of electron-electron interactions is not visible. Thus, the wavepacket remains essentially unsplit and unpolarized ($\Delta \simeq 0$) and the only significant influence is on the average velocity $(v_{+} + v_{-})/2$, from Eq.~(\ref{eq:v_janak}) [see Fig.~\ref{fig:ElectronsVsHoles}(a)]. In contrast, a large splitting is found for $n=3$ where $\Delta,\Delta_0$ can reach a large fraction of the traveling distance. Since spin-polarized ballistic transport is observed in two-dimensional hole systems on $\mu$m-scales (e.g., in spin-focusing experiments\cite{Rokhinson2004,ChesiQPC}) and the typical wavepacket traveling time  is $1 - 50$ ps (depending on the density), Fig.~\ref{fig:SeperationFixedDistance} suggests that a direct optical imaging of the wavepacket separation should be within reach of femtosecond-resolved Faraday rotation measurements.\cite{Crooker1996,Kikkawa1997}

As for electron-electron interaction effects, we see in Fig.~\ref{fig:SeperationFixedDistance} that a visible difference between $\Delta$ and $\Delta_0$ exists. The difference is quite small, due to the fact that we are considering here the weak-coupling limit, and all the interaction corrections are proportional to $r_s < 1$. $\Delta_0$ is modified here by $\sim 2-3\%$  and we collect some representative values in Table~\ref{tab:Values2}. It can be seen in Table~\ref{tab:Values2} that,  while $\Delta_0$ does not change with $r_s$, the interaction correction $\Delta -\Delta_0$ grows at lower densities (see Table~\ref{tab:Values2}). Since experiments on hole systems can reach values as large as $r_s = 6 - 12 $,\cite{Winkler2005,Chiu2011} it is reasonable to expect significant effects from $\delta v_\pm$ in this low-density regime. As a reference, in electron systems, $\delta v$ changes from $\sim 5 \%$ to $-30 \%$ for $r_s \sim 1$ to $6$.\cite{Tan2005,Drummond2009} 

It might be surprising to see $\Delta < \Delta_0$ in Fig.~\ref{fig:SeperationFixedDistance}. This is due to the interplay of two competing interaction effects. It is true that, as discussed already, the difference between $v_{\pm}$ is enhanced by $\delta v_\pm$. The two spin components therefore split faster in the interacting case. However, the situation shown in Fig.~\ref{fig:SeperationFixedDistance} is distinct from that shown in Fig.~\ref{fig:ElectronsVsHoles} since a constant traveling distance (and not time $\tau$) is assumed. At high density the mean velocity $(v_{+}+v_{-})/2$ is greater for the interacting gas ($\delta v >0$), which allows the wavepackets less time to separate. As it turns out, the latter effect is dominant in Fig.~\ref{fig:SeperationFixedDistance}.

Interestingly, the sign of $\delta v$ changes at low density.\cite{TheBook} This would imply a cooperative effect of $\delta v_\pm$ and $\delta v$ on $\Delta$ if $\delta v_\pm$ does not change sign. Unfortunately, to infer the behavior of $\delta v_\pm$ at low density requires a much more sophisticated approach. 

%===================================Seperation of wavepackets at fixed distance Table.
\begin{table}
\begin{center}
    \begin{tabular}{ | l| c | c | c | c | }
    \hline
    SOI    &   $ g$    &  $r_s$    &     $\Delta_0$ (nm)     &     $\Delta-\Delta_0$ (nm)            \\ \hline
    n=1    &   0.05    &  0.5      &     $ 0$                  &     0.001       \\ \hline    
    n=3    &    0.1    &  0.1      &     $ 203$                &    -4.7                          \\ \hline    
    n=3    &    0.1    &  0.3      &     $  203 $              &   -6.2                          \\
    \hline
    \end{tabular}
    \caption{ \label{tab:Values2} Separation of wavepackets after $1~\mu {\rm m}$ drift distance for electrons ($n=1$) and holes ($n=3$). The non-interacting value and the electron-electron interaction correction are listed, see Eq.~(\ref{eq:Delta}) and Fig.~\ref{fig:SeperationFixedDistance}. Typical values for $g$ ($\sim \chi$) are $\lesssim 0.05$ for electrons\cite{Nitta1997,WinklerBook} and $\lesssim 0.2$ for holes. \cite{Winkler2002,WinklerBook} We have assumed in this table that typical values of $g$ are independent of $r_s$ and have used high-density values $r_s<1$.}
\end{center}
\end{table}
%=======================================================================================

%=======================================================================================
\section{Conclusion}\label{sec:conclusion}
%=======================================================================================

We have presented a theory of the quasiparticle group velocity at high density, in the presence of SOI of different types. Contrasting the behavior of electron and hole systems, we find several intriguing differences. We have shown explicitly that the lowest-order cancellations of SOI effects occur only for the electronic case, when the SOI is approximately linear in momentum (e.g., a strong Rashba or Dresslhaus SOI is present). On the other hand, SOI terms non-linear in ${\bf p}$ are often dominant in hole systems.\cite{Winkler2000,Winkler2002,ChesiQPC} Thus, larger effects of the SOI and a non-trivial interplay with electron-electron interactions are expected for holes on general grounds.

As an important motivation for future theoretical studies, hole liquids can be realized in the laboratory with strong SOI and large values of $r_s$. For example, the spin-subband population difference at zero field is of order $15-20\%$ in Ref.~\onlinecite{Winkler2002}, with $n_s \simeq 2-4 \times 10^{10}~{\rm cm}^{-2}$. With a hole effective mass $m \simeq 0.2 m_0$ in GaAs these densities correspond to $r_s \simeq 9 - 12$. For electrons, materials with strong SOI typically have small effective masses, which results in much lower values of $r_s$. A diluted electron liquid with $n_s \simeq 2 \times  10^{10}~{\rm cm}^{-2}$ gives $r_s \simeq 1.2$, using InAs parameters ($m=0.023 m_0$).

Discussing the large-$r_s$ regime of holes would require extending many-body perturbation theory\cite{Saraga2005, Nechaev2009, Nechaev2010} or Monte Carlo\cite{Ambrosetti2009} approaches, so far only applied to linear SOI. The high-density regime studied here would represent a well-controlled limit of these theories for the quasiparticle dispersion. In addition to being relevant for transport measurements of the effective mass, the significant difference in group velocity at the Fermi surface of the two spin branches could also be addressed by Raman scattering experiments, demonstrated for electron systems in Ref.~\onlinecite{Jusserand1992}, or via time-resolved Faraday-rotation detection of the spin-polarization.\cite{Crooker1996,Kikkawa1997} A similar discussion should hold with $n=2,3$ for other physical observables and many-body effects. For example, studying the compressibility\cite{TheBook, Eisenstein1992} in the presence of SOI and extending the $n=1$ discussion of the quasiparticle lifetime\cite{Saraga2005, Nechaev2009, Nechaev2010} would also be topics of interest. Finally, the problem of including in our framework a more general form of SOI than Eq.~(\ref{eq:H0}) is clearly of practical relevance. However, as discussed in Appendix~\ref{app:gneral_arg}, we expect that our qualitative picture on the different role of linear and non-linear SOI remains valid also in this more general situation. 

\acknowledgments
We thank J. P. Eisenstein and D. L. Maslov for useful discussions and acknowledge financial support from CIFAR, NSERC, and FQRNT.

\appendix

%=======================================================================================
\section{General spin-orbit coupling \label{app:gneral_arg}}
%=======================================================================================

A general SOI contains terms with linear and non-linear dependence on momentum, and does not necessarily have the isotropic form assumed in Eq.~(\ref{eq:H0}). For definiteness, we suppose that there is no magnetic field, so that quadratic terms are not present:
\begin{equation}\label{eq:H_general_13}
H= \sum_i \frac{p_i^2}{2m} +H^{\rm so}_1 + H^{\rm so}_3 + H_{\rm el-el},
\end{equation}
where $H^{\rm so}_1$, $H^{\rm so}_3$, and $H_{\rm el-el}$ give, respectively, the linear-in-momentum spin-orbit, the cubic-in-momentum spin-orbit, and electron-electron interactions. To show that $H^{\rm so}_1$ always has a small effect, we consider the unitary transformation $H'= e^{- S} H e^{S}$ with $S$ defined by $[S,\sum_i p_i^2/2m]=H^{\rm so}_{1}$, giving
\begin{equation}\label{eq:transformed_H}
H' \simeq  \frac{p^2}{2m} + H^{\rm so}_3 + H_{\rm el-el} - [S, H^{\rm so}_1 + H^{\rm so}_3].
\end{equation}
In the same transformed frame, the velocity operator of electron $i$ is given by ${\bf v}'_i =  e^{- S} (\partial H/\partial{\bf p}_i) e^{S}$, which yields 
\begin{equation}\label{eq:transformed_v}
{\bf v}' _i \simeq  \frac{{\bf p}_i}{m} + \frac{\partial{H^{\rm so}_3}}{\partial{\bf p}_i} - [S, \frac{\partial{H^{\rm so}_1}}{\partial{\bf p}_i} + \frac{\partial{H^{\rm so}_3}}{\partial{\bf p}_i}].
\end{equation}
In deriving Eq.~(\ref{eq:transformed_H}) we have used the fact that $[S,H_{\rm el-el}]=0$. For example, for an isotropic SOI as in Eq.~(\ref{eq:H0}),
\begin{equation}
S= i \frac{m\gamma}{\hbar}\sum_i (x_i\sigma_{y,i} - y_i \sigma_{x,i}).
\end{equation}
The property $[S,H_{\rm el-el}]=0$ is valid for a general linear-in-momentum SOI, including a combination of Rashba and Dresselhaus SOI.\cite{Aleiner2001} However, the same identity $[S, H_{\rm el-el}]=0$ does not hold for a transformation with $[S,\sum_i p_i^2/2m]= H^{\rm so}_{3}$, i.e., a transformation that is designed to remove the non-linear component from the non-interacting Hamiltonian. In writing Eq.~(\ref{eq:transformed_v}), we have used $[S,{\bf p}_i/m]= \partial{H^{\rm so}_1}/\partial{\bf p}_i $, which implies a cancellation of the $H^{\rm so}_1$ contribution to ${\bf v}'_i$ to lowest order. Again, this cancellation is only valid for linear-in-momentum SOI.

By introducing dimensionless couplings $g_{1,3}$ associated with $ H^{\rm so}_{1,3}$, in direct analogy with Eq.~(\ref{eq:g_def}), we see that the commutators in Eqs.~(\ref{eq:transformed_H}) and (\ref{eq:transformed_v}) are of quadratic or bilinear order in the couplings ($\sim g_1^2$ and $\sim g_1 g_3$). This indicates on general grounds that $H^{\rm so}_3$ has the largest effect on the quasiparticle velocity if $g_1 \lesssim g_3 \ll 1$. In this case, if we are interested in lowest-order effects, we can neglect both anticommutators in Eqs.~(\ref{eq:transformed_H}) and (\ref{eq:transformed_v}), which is equivalent to neglecting $H^{\rm so}_1$ in the original Hamiltonian (\ref{eq:H_general_13}). Thus, to leading order all results we report for the quasiparticle velocities due to a pure cubic-in-momentum spin-orbit interaction also apply in the case of a mixed linear-plus-cubic spin-orbit interaction (with the caveat that we consider only the isotropic form of cubic SOI).

%=======================================================================================
\section{Derivation of Eq.~(\ref{chi_interacting}) \label{app:occupation}}
%=======================================================================================

As discussed in the text, at high density we can neglect the $\sqrt{2}r_s$ in the integrand of Eq.~(\ref{eq:Gamma-n-app}) (second line). For small $g$, the value of $\chi$ is also small and we can perform an expansion of the exchange integral. First notice that the constant term at $\chi=0$ is missing, because the integration limits simply become $y_\pm =1$ and the integrand vanishes upon the summation on $\sigma,\sigma'$. Therefore we only need to compute the linear term in $\chi$:
\begin{eqnarray} \label{eq:dchi_exchange}
&&\left[\frac{\partial}{\partial\chi} \sum_{\sigma \sigma'}  \int_0^{2\pi}\frac{d\theta}{2\pi} \int_{0}^{y_{\sigma'}} 
\frac{(\sigma+\sigma' \cos{n\theta})y dy}{\sqrt{y^2+y_{\sigma}^{2} - 2y y_{\sigma} \cos{\theta}}}\right]_{\chi=0} \nonumber \\
= && \int_0^{2\pi}\frac{d\theta}{2\pi} \left[\int_{0}^{1}\frac{2(1-y \cos{\theta})y dy}{(1+y^2-2y \cos{\theta})^\frac{3}{2}}
-\frac{\cos{n\theta}}{\sin{\frac{\theta}{2}}} \right],
\end{eqnarray}
and after evaluating the $dy$ integral in the square parenthesis, Eq.~(\ref{eq:dchi_exchange}) gives
\begin{eqnarray}\label{appAintegral}
\int_0^{2\pi} && \frac{d\theta}{2\pi} \Bigg[ 2 +\frac{2\cos{\theta}-\cos n\theta-1}{\sin{\frac{\theta}{2}}} \nonumber \\
 -2 && \, \ln\left(1+\frac{1}{\sin{\frac{\theta}{2}}} \right)\cos{\theta} \Bigg]=\frac{4}{\pi} \sum_{j=0}^n \frac{1}{2j-1}.
\end{eqnarray}
Equation~(\ref{chi_interacting}) is then easily obtained from Eq.~(\ref{eq:Gamma-n-app}) by neglecting all the cubic terms in the small parameters $g,\chi,r_s$ [e.g., the left side of Eq.~(\ref{eq:Gamma-n-app}) is $(y_+^2+y_-^2)g= 2g+O(g\chi^2)$].

%=======================================================================================
\section{Small $r_s,g$ expansions \label{app:expansions}}
%=======================================================================================

We give in this appendix some details on the expansions of Eq.~(\ref{eq:groupvelocity_explicit}):
\begin{equation}
\frac{v_{F \pm}}{v_{F}} =   y_\pm \pm \frac{g}{2} n y_\pm ^{n-1} +\frac{\sqrt{2}r_s}{16\pi}(I_{1}+I_{2}+I_3),
\end{equation}
where we have split $L_2$ in its two contributions ($I_{1,2}$ refer to $\sigma=\pm$) and $I_3$ corresponds to $L_3$:
\begin{eqnarray}
I_1 &=& \int_0^{2\pi} d \theta \frac{\sqrt{2} y_\pm \cos{\theta}(1+\cos{n \theta})}{\sqrt{2} y_\pm  \sin{\theta/2} + r_s },\label{eq:I1}\\
I_2 &=& \int_0^{2\pi} d \theta \frac{\sqrt{2} y_\mp \cos{\theta} (1-\cos{n\theta})}{\sqrt{1-y_+y_- \cos{
\theta}} + r_s},\label{eq:I2}\\
I_3 &=& \pm \int_0^{2\pi} d \theta \int_{y_{+}}^{y_{-}} dy \frac{2n\sin{n\theta}\sin{\theta}}{\sqrt{y_{\pm}^2 +y^2 - 2 y y_{\pm} \cos{\theta}} + r_s \sqrt{2}},\label{eq:I3}
\end{eqnarray}
Notice that these integrals have an explicit dependence on $r_s$ and $y_\pm  =\sqrt{1\mp \chi}$. So, it is easier to perform first the expansion in the two small parameters $r_s,\chi$. The final results in the main text are given in terms of the physical couplings of the hamiltonian: $r_s$ and $g$. Those final expression are easily obtained by substituting the value of $\chi$ in terms of $r_s$ and $g$ ($\chi \simeq g$ in first approximation). 

The first integral, Eq.~(\ref{eq:I1}), can be evaluated exactly. In particular for $n=1$ we obtain
\begin{eqnarray}\label{eq:I1_expansion}
I_1 = \, && -\frac{40}{3}+8\pi \delta(1-\delta^2)+16\delta^2 \nonumber \\
&& +8 (1-3\delta^2 +2\delta^4)\frac{\tanh^{-1}\sqrt{1-\delta^2}}{\sqrt{1-\delta^2}}
\end{eqnarray}
where $\delta=r_s/\sqrt{2}y_\pm$. This expression can then be easily expanded in $r_s,\chi$ and an analogous procedure is followed for $n=2,3$. To lowest-order in $r_s$, we can set $r_s=0$ in $I_2$ and $I_3$. Similarly to the $I_1$ angular integral above, the $d\theta$ integrals of $I_2$ and $I_3$ at $r_s=0$ can be computed analytically for $n=1,2,3$. For $I_2$ this yields directly the desired function of $\chi$. For $I_{3}$ we still need to perform a last integration in $dy$. Since the integration region is of size $\sim\chi$ around $y=1$, we can expand the integrand in the small parameter $(y-1)$ and perform the integration in $dy$ order-by-order, which allows us to extract the leading terms of the expansion in $\chi$. For $n=1$ all this gives
\begin{eqnarray}
\delta I_1 &\simeq & \mp 4 \chi - 2 \chi^2 \mp \frac43 \chi^3,\\
\delta I_2 &\simeq & \mp \frac43 \chi + \chi^2 \left(\ln\frac{\chi}{8}+\frac{13}{6}\right) \pm  \frac{\chi^3}{2} \left(\ln\frac{\chi}{8}+ \frac32\right),\\
\delta I_3 &\simeq & \pm \frac{16}{3}\chi + \frac43 \chi^2 \pm \chi^3\left(\frac23 \ln\frac{\chi}{16} + \frac{389}{120}\right) ,
\end{eqnarray}
where only the corrections $\delta I_{\alpha}=I_\alpha(\chi)-I_{\alpha}(\chi=0)$ are listed, since terms independent on $\chi$ simply give the small $r_s$ expansion of the well known Eq.~(\ref{eq:v_janak}). Here, terms of order  $O(r_s \chi, \chi^4)$ are omitted, while it is interesting to keep the $O(\chi^3)$ terms, since they give the leading spin splitting. Indeed, it is easily checked that the linear terms cancel
\begin{equation}
\sum_{\alpha =1}^3 \delta I_\alpha \simeq 
\, \chi^2 \left(\frac{3}{2} + \ln\frac{\chi}{8}\right) \pm  \frac{\chi^3}{6}\left( \ln\frac{\chi^7}{8^5} + \frac{319}{20}\right),
\end{equation}
which immediately leads to Eq.~(\ref{eq:n1_v_final}). 

For $n=2,3$ we can proceed in a similar way. The spin splitting appears now already to linear order in $\chi$. By keeping the first subleading correction in $\chi$ we have for $n=2$:
\begin{eqnarray}
\delta I_1 &\simeq & \mp 4 \chi - 2 \chi^2,\\
\delta I_2 &\simeq & \pm \frac{16}{15} \chi + \chi^2 \left(4 \ln{\frac{\chi}{8}}+\frac{134}{15}\right),\\
\delta I_3 &\simeq & \pm \frac{64}{15} \chi + \frac{16}{15}\chi^2,
\end{eqnarray}
and for $n=3$:
\begin{eqnarray}
\delta I_1 &\simeq & \mp 4 \chi - 2 \chi^2,\\
\delta I_2 &\simeq & \pm \frac{212}{105} \chi + \chi^2 \left( 9 \ln{\frac{\chi}{8}}+\frac{899}{42}\right),\\
\delta I_3 &\simeq & \pm \frac{144}{35} \chi + \frac{36}{35}\chi^2.
\end{eqnarray}
The final results are for $n=2$:
\begin{equation}\label{eq:Isum_2}
\sum_{\alpha =1}^3 \delta I_\alpha \simeq 
\, \pm \frac{4}{3} \chi + 4 \chi^2 \left( \ln\frac{\chi}{8}+2 \right),
\end{equation}
and for $n=3$:
 \begin{equation}\label{eq:Isum_3}
\sum_{\alpha =1}^3 \delta I_\alpha \simeq 
\, \pm \frac{32}{15} \chi + \chi^2 \left(9\ln\frac{\chi}{8}+\frac{613}{30} \right).
\end{equation}
From Eqs.~(\ref{eq:Isum_2}) and (\ref{eq:Isum_3}) we immediately obtain Eqs.~(\ref{eq:L23_2}) and (\ref{eq:L23_3}).

%=======================================================================================

%=======================================================================================

\begin{thebibliography}{100}
%=======================================================================================
\bibitem{TheBook} F. Giuliani and G. Vignale, \emph{Quantum Theory of the Electron
Liquid} (Cambridge University Press, Cambridge, 2005).
\bibitem{WinklerBook} R. Winkler, \emph{Spin-Orbit Coupling Effects in Two-Dimensional
Electron and Hole Systems} (Springer, Berlin, 2003).
\bibitem{Datta1989} S. Datta and B. Das, Appl. Phys. Lett. {\bf 56}, 665 (1989).
\bibitem{Nitta1997} J. Nitta, T. Akazaki, H. Takayanagi, and T. Enoki, Phys. Rev. Lett. {\bf 78}, 1335 (1997).
\bibitem{Johnson2009}H. C. Koo, J. H. Kwon, J. Eom, J. Chang, S. H. Han, and M. Johnson, Science {\bf 325}, 1515 (2009).   
\bibitem{Chen1999} G.-H. Chen and M. E. Raikh, Phys. Rev. B {\bf 60}, 4826 (1999).
\bibitem{Saraga2005} D. S. Saraga and D. Loss,, Phys. Rev. B {\bf 72}, 195319 (2005).
\bibitem{Nechaev2009} I. A. Nechaev, M. F. Jensen, E. D. L. Rienks, V. M. Silkin, P. M. Echenique, E. V. Chulkov, and P. Hofmann, Phys. Rev. B {\bf 80}, 113402 (2009).
\bibitem{Nechaev2010} I. A. Nechaev, P. M. Echenique, and E. V. Chulkov, Phys. Rev. B {\bf 81}, 195112 (2010).
\bibitem{Agarwal2011} A. Agarwal, S. Chesi, T. Jungwirth, J. Sinova, G. Vignale, and M. Polini, Phys. Rev. B {\bf 83}, 115135 (2011).
\bibitem{Bychov1984a} Y. A. Bychkov and E. Rashba, JETP Lett. {\bf 39}, 78 (1984).
\bibitem{Bychov1984b} Y. A. Bychkov and E. Rashba, J. Phys. C {\bf 17}, 6039 (1984).
\bibitem{Dresselhaus1955} G. Dresselhaus, Phys. Rev. {\bf 100}, 580 (1955).
\bibitem{Dyakonov1986} M. I. D'yakonov and V.Yu. Kachorovskii, Sov. Phys. Semicond. {\bf 20}, 110 (1986). 
\bibitem{Berman2010} D. H. Berman and M. E. Flatt\'e, Phys. Rev. Lett. {\bf 105}, 157202 (2010).
\bibitem{ChesiExact} S. Chesi and G. F. Giuliani, Phys. Rev. B {\bf 83}, 235308 (2011).
\bibitem{Abedinpour2010} S. H. Abedinpour, G. Vignale, and I. V. Tokatly, Phys. Rev. B {\bf 81}, 125123 (2010). 
\bibitem{Ambrosetti2009} A. Ambrosetti, F. Pederiva, E. Lipparini, and S. Gandolfi, Phys. Rev. B {\bf 80}, 125306 (2009).
\bibitem{ChesiHighDens} S. Chesi and G. F. Giuliani, Phys. Rev. B {\bf 83}, 235309 (2011). 
\bibitem{ChesiHFstates} G. F. Giuliani and S. Chesi, Proceedings of \emph{Highlights in the
Quantum Theory of Condensed Matter} (Edizioni della Normale, Pisa, 2005); S. Chesi, G. Simion, and G. F. Giuliani, e-print
arXiv:cond-mat/0702060; S. Chesi, Ph.D. thesis, Purdue University, 2007.
\bibitem{Juri2008} L. O. Juri and P. I. Tamborenea, Phys. Rev. B {\bf 77}, 233310 (2008).
\bibitem{Zak2010} R. A. \.Zak, D. L. Maslov, and D. Loss, Phys. Rev. B {\bf 82}, 115415 (2010).
\bibitem{Zak2011} R. A. \.Zak, D. L. Maslov, and D. Loss, arXiv:1112.4786.
\bibitem{Aleiner2001}  I. L. Aleiner and V. I. Falko, Phys. Rev. Lett. {\bf 87}, 256801 (2001). 
\bibitem{Coish2006} W. A. Coish, V. N. Golovach, J. C. Egues, and D. Loss, phys. stat. sol (b) {\bf 243}, 3658 (2006).
\bibitem{Winkler2000} R. Winkler, Phys. Rev. B {\bf 62}, 4245 (2000).
\bibitem{Winkler2002} R. Winkler, H. Noh, E. Tutuc, and M. Shayegan, Phys. Rev. B {\bf 65}, 155303 (2002).
\bibitem{Rokhinson2004} L. P. Rokhinson, V. Larkina, Y. B. Lyanda-Geller, L. N. Pfeiffer,
and K. W. West, Phys. Rev. Lett. {\bf 93}, 146601 (2004).
\bibitem{ChesiQPC} S. Chesi, G. F. Giuliani, L. P. Rokhinson, L. N. Pfeiffer, and K. W. West, Phys. Rev. Lett. {\bf 106}, 236601 (2011).
\bibitem{Bulaev2007} D. V. Bulaev and D. Loss, Phys. Rev. Lett. {\bf 98}, 097202 (2007). 
\bibitem{Chesi07April} S. Chesi and G. F. Giuliani, Phys. Rev. B {\bf 75}, 155305 (2007).
\bibitem{Rashba1988} E. I. Rashba and E. Ya. Sherman, Phys. Lett. A {\bf 129}, 175 (1988).
\bibitem{Winkler2005} R. Winkler, E. Tutuc, S. J. Papadakis, S. Melinte, M. Shayegan, D. Wasserman, and S. A. Lyon, Phys. Rev. B {\bf 72}, 195321 (2005).
\bibitem{Chiu2011} Y. T. Chiu, M. Padmanabhan, T. Gokmen, J. Shabani, E. Tutuc, M. Shayegan, and R. Winkler, e-print arXiv:1106.4608.
\bibitem{Crooker1996} S. A. Crooker, J. J. Baumberg, F. Flack, N. Samarth, and D. D. Awschalom, Phys. Rev. Lett. {\bf 77}, 2814 (1996).
\bibitem{Kikkawa1997} J. M. Kikkawa and D. D. Awschalom, Phys. Rev. Lett. {\bf 80}, 4313 (1998).
\bibitem{Tan2005} Y.-W. Tan, J. Zhu, H. L. Stormer, L. N. Pfeiffer, K. W. Baldwin, and K. W. West, Phys. Rev. Lett. {\bf 94}, 016405 (2005).
\bibitem{Drummond2009} N. D. Drummond and R. J. Needs, Phys. Rev. B {\bf 80}, 245104 (2009).
\bibitem{Janak1969} J. F. Janak, Phys. Rev. {\bf 178}, 1416 (1969).
\bibitem{comment_gen_chi} To deal with extremely low densities at which only one spin band is occupied and the occupation is a ring in momentum space, it is also useful to generalize $\chi$ to values larger than one.\cite{ChesiHighDens} However, we do not consider this low-density regime in this paper. 
\bibitem{Schliemann2006} J. Schliemann, Phys. Rev. B {\bf 74}, 045214 (2006).
\bibitem{comment_variational} The method of Ref.~\onlinecite{Chesi07April} is slightly preferable, in that it leads to variational wavefunctions with a lower total energy. 
\bibitem{Jusserand1992} B. Jusserand, D. Richards, H. Peric, and B. Etienne, Phys. Rev. Lett. {\bf 69}, 848 (1992). 
\bibitem{Eisenstein1992} J. P. Eisenstein, L. N. Pfeiffer, and K. W. West, Phys. Rev. Lett. {\bf 68}, 674 (1992).
\end{thebibliography}
\end{document}